\documentclass[letterpaper,10pt,conference]{IEEEtran}

\IEEEoverridecommandlockouts

\usepackage{lipsum}
\usepackage{graphics,graphicx}
\usepackage{fancyhdr,lipsum}

\makeatletter
 \let\old@ps@headings\ps@headings
 \let\old@ps@IEEEtitlepagestyle\ps@IEEEtitlepagestyle
 \def\confheader#1{%
 \def\ps@headings{%
 \old@ps@headings%
 \def\@oddhead{\strut\hfill#1\hfill\strut}%
 \def\@evenhead{\strut\hfill#1\hfill\strut}%
 }%
 \def\ps@IEEEtitlepagestyle{%
 \old@ps@IEEEtitlepagestyle%
 \def\@oddhead{\strut\hfill#1\hfill\strut}%
 \def\@evenhead{\strut\hfill#1\hfill\strut}%
 }%
 \ps@headings%
 }
 \makeatother

\confheader{%
2023 IEEE World AI IoT Congress (AIIoT)
 }

 \usepackage[pscoord]{eso-pic}
\newcommand{\placetextbox}[3]{
 \setbox0=\hbox{#3}
 \AddToShipoutPictureFG*{ \put(\LenToUnit{#1\paperwidth},\LenToUnit{#2\paperheight}){\vtop{{\null}\makebox[0pt][c]{#3}}}
 }
 }
 \placetextbox{.23}{0.055}{\small{979-8-3503-3761-7/23/\$31.00~\copyright 2023 IEEE}}

\begin{document}

\title{Implementing Man-in-the-Middle Attack to Investigate Network Vulnerabilities in Smart Grid Test-bed}

\author{
  \IEEEauthorblockN{Shampa Banik}%
  \textit{IEEE Student Member}\\
 \IEEEauthorblockA{\textit{Dept. of Computer Science} \\
\textit{Tennessee Tech University}\\
Tennessee, USA \\
sbanik42@tntech.edu}
   
   \and
   
\IEEEauthorblockN{Trapa Banik}%
  \textit{IEEE Student Member}\\
 \IEEEauthorblockA{\textit{Dept. of ECE}\\
\textit{Tennessee Tech University}\\
Tennessee, USA \\tbanik42@tntech.edu}

\and

\IEEEauthorblockN{S M Mostaq Hossain}%
  \textit{IEEE Student Member}\\
 \IEEEauthorblockA{\textit{Dept. of Computer Science} \\
\textit{Tennessee Tech University}\\
Tennessee, USA \\shossain42@tntech.edu}

\and

\IEEEauthorblockN{Sohag Kumar Saha}%
  \textit{IEEE Student Member}\\
  \IEEEauthorblockA{\textit{Dept. of ECE} \\
\textit{Tennessee Tech University}\\
Tennessee, USA \\
ssaha42@tntech.edu}

}





\maketitle


\pagestyle{fancy}
\fancyfoot[C]{\thepage}

\maketitle

\begin{abstract}
The smart-grid introduces several new data-gathering, communication, and information-sharing capabilities into the electrical system, as well as additional privacy threats, vulnerabilities, and cyber attacks. In this study, Modbus is regarded as one of the most prevalent interfaces for control systems in power plants. Modern control interfaces are vulnerable to cyber-attacks, posing a risk to the entire energy infrastructure. In order to strengthen resistance to cyber attacks, this study introduces a test bed for cyber-physical systems that operate in real-time. To investigate the network vulnerabilities of smart power grids, Modbus protocol has been examined combining a real-time power system simulator with a communication system simulator and the effects of the system presented and analyzed. The goal is to detect the vulnerability in Modbus protocol and perform the Man-in-the-middle attack with its impact on the system. This proposed testbed can be evaluated as a research model for vulnerability assessment as well as a tool for evaluating cyber attacks and enquire into any detection mechanism for safeguarding and defending smart grid systems from a variety of cyberattacks. We present here the preliminary findings on using the testbed to identify a particular MiTM attack and the effects on system performance. Finally, we suggest a cyber security strategy as a solution to address such network vulnerabilities and deploy appropriate countermeasures. 

\end{abstract}

\section{INTRODUCTION}
\IEEEPARstart{S}{mart grid} is a new paradigm in the energy industry that has replaced the conventional electric grid. The smart grid harnesses the power of information technology to efficiently meet environmental criteria by facilitating the integration of green technologies and intelligently delivering electricity to clients via two-way communication. The addition of information and communications technology (ICT) services to the conventional electrical grid represents a technical advancement. It makes use of information interchange in both directions to construct an automated and widely dispersed system with extra features like real-time control, operational efficiency, grid resilience, and increased integration of renewable technology, all of which minimize carbon footprint.

Although SG offers numerous advantages, it also poses significant security risks due to the fact that it combines disparate communication networks, such as those for the Internet of Things (IoT), industrial devices, wireless components, and Wireless Sensor Networks (WSN), which are vulnerable to a variety of security threats. Any disruptions in power production could damage the smart grid's reliability and have substantial socioeconomic implications. In addition, because smart grid systems communicate sensitive information, loss or manipulation of this information could compromise user privacy. These network vulnerabilities have made smart grid the primary target of attackers, attracting the interest of the government, corporations, and academia. Industrial communication protocols are utilized by smart grids and Supervisory Control and Data Acquisition (SCADA) networks to facilitate the operation of Industrial Control Systems (ICS)\cite{R1}. However, widely utilized protocols such as Modbus and DNP3 lack fundamental security measures, resulting in a multitude of vulnerabilities. Such network vulnerabilities could have a considerable impact on businesses and the general population, especially if attacks target critical infrastructure components such as power plants, water distribution networks, and rail networks. Solar distributed energy resources (DER), where devices are frequently put in client premises and security is difficult to implement, exacerbate the situation. Test-beds are essential for power grid cyber-physical systems research due to a shortage of modeling tools and the need for exact security evaluation. The switching module, which transmits, processes, and analyzes data, is critical but vulnerable. Because communication protocol flaws can be exploited, a smart grid's cybersecurity threats increase.  

In this study, our main focus is to investigate its infrastructure for cyber-attack, one of the most significant network protocols, Modbus\cite{R2}. The basic model of a smart grid system, as depicted in Fig. 1, is based on the interconnection of dependent domains employed for energy production, transmission, and distribution\cite{R3}. 

\begin{figure}[h!]
\centering
\includegraphics[width=1\columnwidth]{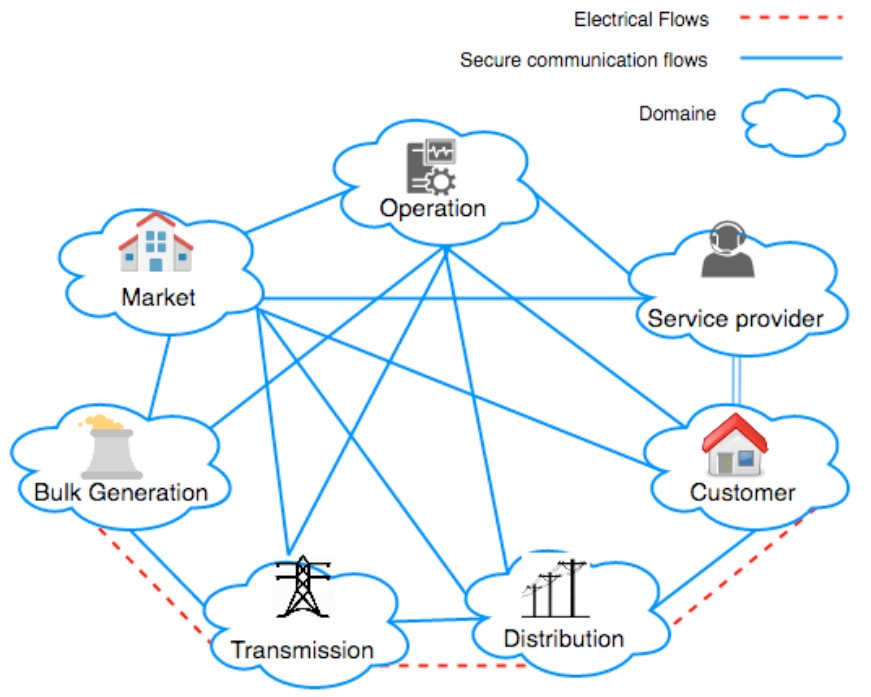}
\caption{Smart Grid’s conceptual model \cite{R17}}
\label{Fig. 1}
\end{figure}

Modbus is the most frequently used protocol in smart grids, although it lacks many current security features in practice is depicted in Fig. 2. Due to the fact that Modbus/TCP lacks an authentication or access control mechanism, possible cyber attackers can launch a Man-in-the-middle (MiTM) attack and acquire unauthorized access.

\begin{figure}[h!]
\centering
\includegraphics[width=1\columnwidth]{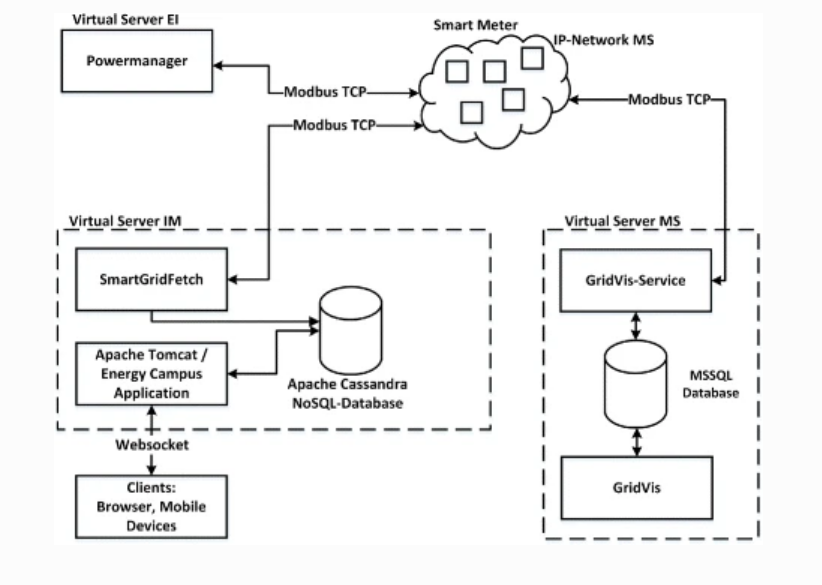}
\caption{Modbus Protocol in Smart Grid \cite{R26}}
\label{Fig. 1}
\end{figure}

In particular, this paper makes two contributions: first, we explore the design and development of the testbed that will simulate the smart grid system by mimicking the communication network between server and client; and second, we exploit the cyber-attack to monitor the system's network vulnerabilities and the effects of the attack\cite{R4}. Metasploit is assisting in this attack. When it comes to testing and utilizing Modbus/TCP, this is one of the most widely recognized tools in the business. With this method, it is feasible to comprehend the actions taken by a hacker to compromise the smart grid's security. The findings of our study will be useful for cyber forensics investigations into the Modbus protocol, including preventative cyber security methods for the smart grid and projections for the future.

The rest of this study is structured as follows. In section II we present the background literature and the research motivation. In section III we discuss the general overview of the Modbus protocol. Section IV is shown the proposed simulated testbed where we examined the attack to penetrate the system by an attacker in the middle of the system along with the experimental demonstration. After that section, V discusses the result and analysis of the implementation. Different mitigation techniques for the Modbus protocol have been outlined in section VI. The final section VII has drawn the conclusion with the future research direction.

\section{RESEARCH OVERVIEW}
To explore deeper into Modbus exploitation in the smart grid many recent pertinent research studies have been reviewed. Figure 2 depicts the detection level, anomaly types, and computing methodology. The authors in \cite{R1} implemented a real-time anomaly detection system that is based on data gathered from smart meters (SM) installed at customers’ homes. P. Huitsing et al. published a thorough theoretical analysis regarding the numerous cyberattacks against the Modbus protocol in \cite{R2}, to be more precise.  In \cite{R4} carried out four different cyberattacks on the Modbus protocol in order to gather relevant traces that might be exploited by machine learning algorithms. To detect intrusions with similar properties to known attacks Neural networks and decision trees were used to classify the traffic generated from this simulated environment. 
Voyiatz is et al. \cite{R5} presented the design and implementation of MTF, a Modbus/TCP Fuzzer. The MTF's testing process includes a reconnaissance phase to aid in mapping the tested device's capabilities and adjusting the attack vectors.
In \cite{R6} the authors described a set of experiments that shows that an anomaly-based change detection algorithm and signature-based Snort threshold module are capable of detecting Modbus flooding attacks. Modbus-related rules were published by T. Morris et al. in \cite{R7}, and they can be used by well-known signature-based IDS like Snort and Suricata. For Modbus/TCP and Modbus over Serial Line networks, researchers developed a set of rules for intrusion detection systems. A vulnerability analysis of the Modbus protocols led to the rules.
 
In light of this, N. Goldenberg and A. Wool provided an appropriate anomaly-based IDS in \cite{R8} that utilizes a Deterministic Finite Automaton (DFA).  Here, a model-based intrusion detection system is based on the fact that Modbus traffic to and from a specific PLC is highly periodic. Because of this, each HMI-PLC channel can be modeled using its own unique deterministic finite automaton (DFA).
 
This is similar to how S. Anton et al. used and assessed different machine learning classification algorithms for identifying Modbus threats in \cite{R9}. To detect fraudulent traffic in a synthetically generated data set of Modbus/TCP communication in a fictitious industrial situation, various machine learning-based anomaly detection methods were applied to an industrial Modbus/TCP data set. In \cite{R10}, the authors proposed an intrusion detection method that uses a honeypot for the Modbus TCP protocol. The proposed approach employs an unsupervised clustering method based on log sequence similarity to cluster attacking behaviors.
The Modbus/TCP Attack-Tree model was implemented to exploit a set of attacks on a CPS testbed \cite{R11}. Replay, Man-in-the-Middle, Denial-of-Service, and Reconnaissance attacks are all included in this model. This serves as a formal risk assessment model that integrates attack time, detection time, and plant hazard generation time.
The authors in paper \cite{R12} provided an analysis of 37 cases and then examined the contribution of the IDPSs in the SG paradigm. By timely detecting or/and preventing potential security violations, these systems can be considered as a secondary defense mechanism, which enhances the cryptographic processes.
The authors in \cite{R13} proposed an IDS that works in two-fold ways. First, they studied and enhanced the cyberattacks provided by the Smod pen-testing tool. Second, they introduce an intrusion detection system (IDS) based on anomalies that could identify intrusions that caused Modbus/TCP denial of service (DoS). 
A case study outlines some countermeasures that may improve security for Modbus/TCP protocol in SCADA systems by examining and testing the existing security countermeasures that are unique to SCADA systems \cite{R14}.
Another study in \cite{R15} presented a model for Organizing information security in Modbus TCP. 

In another review, paper \cite{R16} the authors provided descriptions of several severe cyber-attacks and proposed a cyber-security strategy to detect and counter these attacks. 

In \cite{R18} the authors performed extensive research on the documented and reported network vulnerabilities in the Modbus protocol, and also review various efforts on how to exploit those network vulnerabilities in developing the virtual testbed.

The authors in \cite{R19} presented a set of 17 attacks against SCADA control systems. Reconnaissance, response and measurement injection, command injection, and denial of service are the four categories into which the attacks are divided.

Using reinforcement learning (RL) algorithms, the paper \cite{R20} presented a survey of detection and mitigation techniques against denial-of-service (DoS) attacks on the Modbus protocol in the smart grid, assessing the efficacy of existing approach and identifying the prospects for future research.

Another approach in \cite{R21} proposed a Neural Network architecture named MODLSTM, which consists of three parts: input preprocessing, feature recording, and traffic classification to detect DoS attacks related to Modbus.
In \cite{R22} a cyber-resilience enhancement framework was proposed with the aim to modernize the traditional electrical grid and provide solutions in the domains of voltage and frequency restoration, cybersecurity, and network Quality of Service. In a controlled setting, a technique for breaching the integrity and authentication of packets was demonstrated in \cite{R23}  using the IEEE Synchrophasor Protocol, but the same strategy may be applied to any other protocol that lacks the necessary overhead to guarantee packet integrity and authenticity. The paper \cite{R24} presented an overview of such cyber vulnerabilities, practical limitations of modern power systems, relevant prevention measures, and a case study in detecting network vulnerabilities of modern power systems.

\section{Modbus Protocol}

The Modbus TCP/IP protocol is widely used in industrial automation control systems. These systems oversee critical processes including the regulation of gas turbines and refineries. The protocol was created without any thought to security decades ago. Modbus comes in three different forms: Modbus ASCII, Modbus RTU, and Modbus/TCP. In the first one, hexadecimal is used to code messages. Even though it is slow, it works great for phone calls and radio links. In the second one, messages are sent over RS232 and are coded in binary. In the third, the masters and slaves talk to each other through IP addresses \cite{R6}. In a SCADA system, Modbus is a master-slave protocol that allows one master, remote terminal unit (RTU), or master terminal unit (MTU), and several slave devices, such as sensors, drivers, and PLCs, to talk to each other and share instructions \cite{R7}. On the one hand, Modbus is used a lot in industrial architecture because it is easy to use and can send raw data without authentication, encryption \cite{R8}.
Modbus/TCP was chosen for these reasons in particular:

\begin{itemize}
    \item Modbus is still widely used in power systems.
    \item Modbus/TCP is simple and easy to implement
    \item Modbus protocol libraries are freely available for utilities to implement smart grid applications. 
\end{itemize}

Fig. 3 shows the application layer Modbus.

\begin{figure}[h!]
\centering
\includegraphics[width=1\columnwidth]{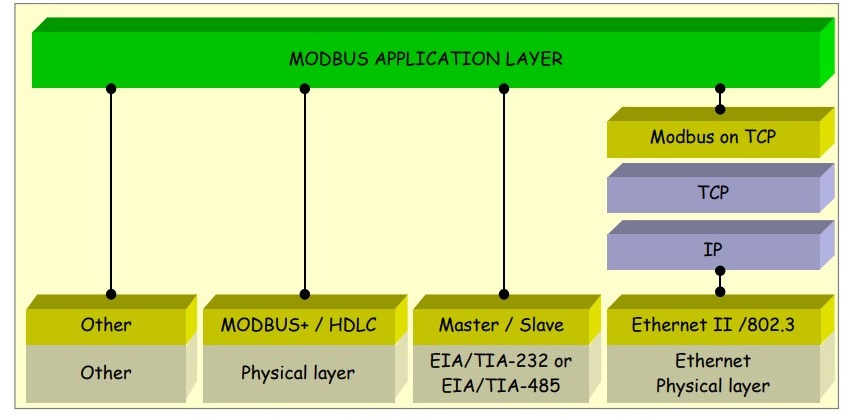}
\caption{Stacks for the application layer: Modbus \cite{R25}}
\label{Fig. 2}
\end{figure}

\subsection{Modbus Cyberattacks}
As depicted in Fig. 4, there are four stages that malicious hackers often go through to attack and take control of the Modbus protocol: reconnaissance, scanning, exploitation, and maintaining access. Reconnaissance is the first phase in which the attacker learns as much as it can about the target. The attacker searches for system network vulnerabilities in the second stage, scanning. These operations seek to locate the opened ports and learn about the services being used by each port, as well as their flaws. He or she attempts to compromise and seize complete control of the target during the exploitation phase. The third phase, which is retaining access, is taken once the attacker has administrative access to the target.
Installing a covert and undetectable program allows the attacker to complete this step and then return quickly to the target machine.
\begin{figure}[h!]
\centering
\includegraphics[width=1\columnwidth]{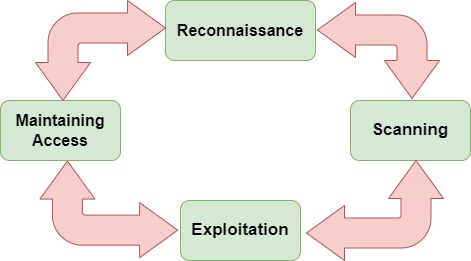}
\caption{Attacking cycle followed by hackers to get control over a system.}
\label{Fig. 3}
\end{figure}

\subsection{What is a man-in-the-middle attack?}
An attack in which the attacker places himself in the path of communication between the user and the system to intercept and change data that is being sent between them.
Attackers use the same techniques in the smart grid to circumvent security measures \cite{R18}. They employ several methods to compromise a certain grid system during each stage. As a result, these actions can be used to categorize attacks.
The types of attacks that occur during each stage are shown in Fig. 4. As one can see, there are many different kinds of attacks that might occur during the exploitation stage. The following steps' malicious actions and assaults are explained.

\section{REAL-TIME TESTBED IMPLEMENTATION}

This paper presents a real-time testbed that is based on the experiment to carry out the MiTM exploitation of the Modbus/TCP protocol. In particular, first, we develop the simulated Modbus environment and test the communication between server and client. Then perform the MiTM attack to get access and detect network vulnerabilities that will further help us to do the injection attack on this protocol. Subsequently, by examining this observation we can come up with ideas for counter-measure against such intrusion as well as such cyber attacks related to Modbus/TCP.

Hence, in short, this testbed is implemented with an aim to address the following rising problems Modbus protocol.

\begin{itemize}
\item Vulnerability analysis
\item Estimating the level of exploitation
\end{itemize}

In order to establish our testbed-based solution model on Modbus attack detection system, components are obtained by using one the following procedure:

\begin{itemize}
\item Building Modbus communication network environment
\item Monitoring data transmission through the Modbus
\item Run the attack as MiTM
\item Intercept the resource data
\item Manipulating the resource data
\item Exploiting the system
\end{itemize}

\subsection{Experimental Setup}

\subsubsection{Modbus Master and Slave Communication Environment Setup}

Numerous Modbus Slave Simulators are available. In this project, we'll utilize ModbusPal, a Java-based Modbus Slave emulator that works with any operating system that supports Java. This offers a GUI interface and makes it simple to create several Modbus slaves.

It emulates a Modbus Gateway with the virtual machine's IP address while running ModbusPal on just one virtual machine (VM1) (accepts TCP connections on port 502). Each Slave that we build using ModbusPal is identifiable by its Unit ID (ranging from 1 to 247).
The network setup is illustrated in Figure 5.
\begin{figure}[h!]
\centering
\includegraphics[width=1\columnwidth]{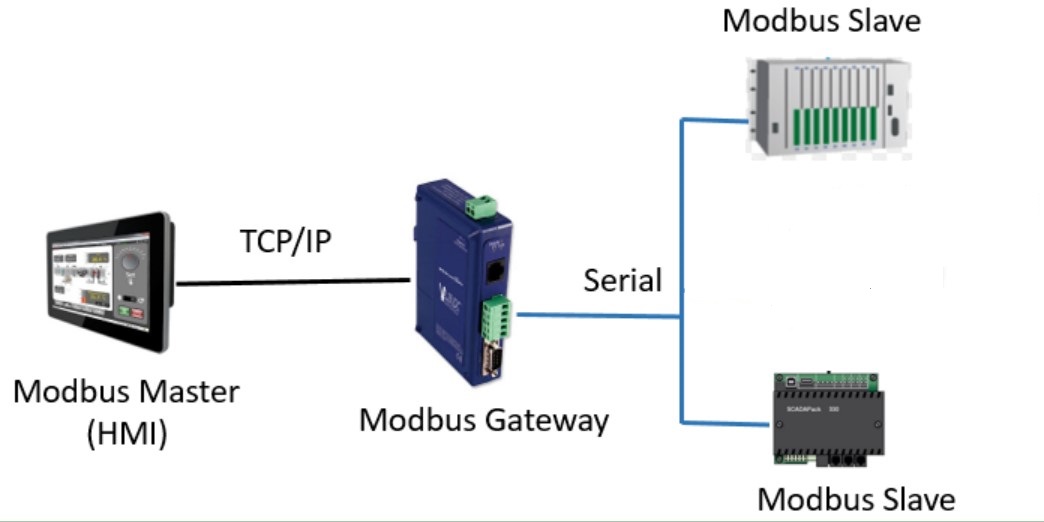}
\caption{Network Setup of Modbus Master and Slaves Using Two Virtual Machines.}
\label{Fig. 4}
\end{figure}

\subsubsection{Communication between Modbus Master and Modbus Slaves}
We use another virtual machine to run the Modbus Master Simulator for master (VM2). The emulated Modbus Master can control/access those Modbus Slaves behind the Modbus Gateway as long as virtual machines 1 and 2 (VM1 and VM2) can establish a connection to each other. Two industrial protocols susceptible to scanning attacks are Modbus and DNP3. Since Modbus/TCP was created more for communication than for security, it can be vulnerable to the Modbus network scanning attack. The packet has been captured to ensure the communication between Modbus master and slave by the means of wireshark in Fig. 6.

\begin{figure}[h!]
\centering
\includegraphics[width=1\columnwidth]{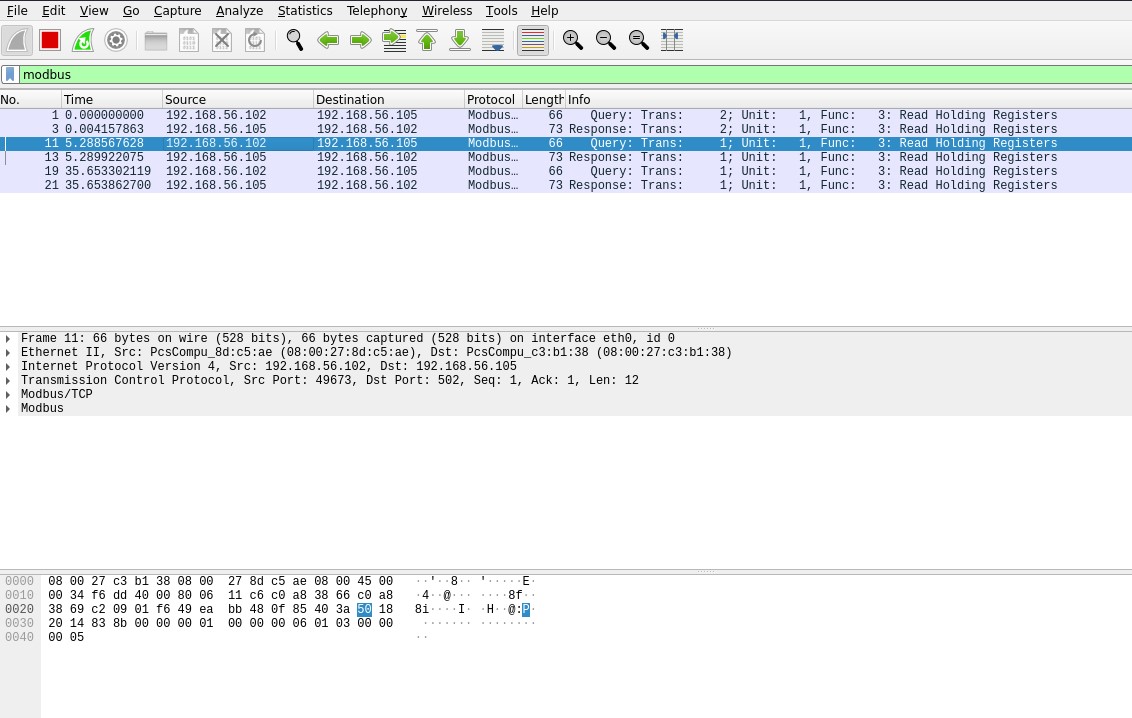}
\caption{Wireshark Capturing Query/Response Packets between Modbus Master and Slave.}
\label{Fig. 5}
\end{figure}

\subsubsection{Using Metasploit to Attack Modbus Slaves}
As the Modbus Master communicates with Slaves in plain text and there is no authentication procedure, an attacker can easily generate the same format of query packets to Modbus Slaves to access/modify Slave's registers/coils, as long as:
The attacker's machine can send packets to Modbus Slave.
The packets sent by the attacker follow Modbus protocol format.
The second requirement can be overcome by using Metasploit which attackers have incorporated Modbus attack modules in Metasploit, thus  Fig. 7. depicts how the attacker intercepts the communication in the middle between the master and client.

\begin{figure}[h!]
\centering
\includegraphics[width=1\columnwidth]{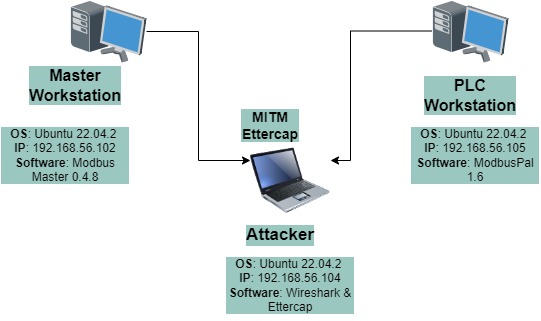}
\caption{MiTM Attack.}
\label{Fig. 6}
\end{figure}

\subsubsection{Intercepting Modbus resource data}
Modbus protocol consists of two types of addresses. 1) Device address 2) Coil and Register Address. The memory address for coil and register of Modbus protocol is shown in Table 1. Two of them storing discrete values 1-bit data called coils and storing 16-bit (int) numerical data values called registers. The table shows there are two read-only entries and two read/write entries.
\begin{figure}[h!]
\centering
\includegraphics[width=1\columnwidth]{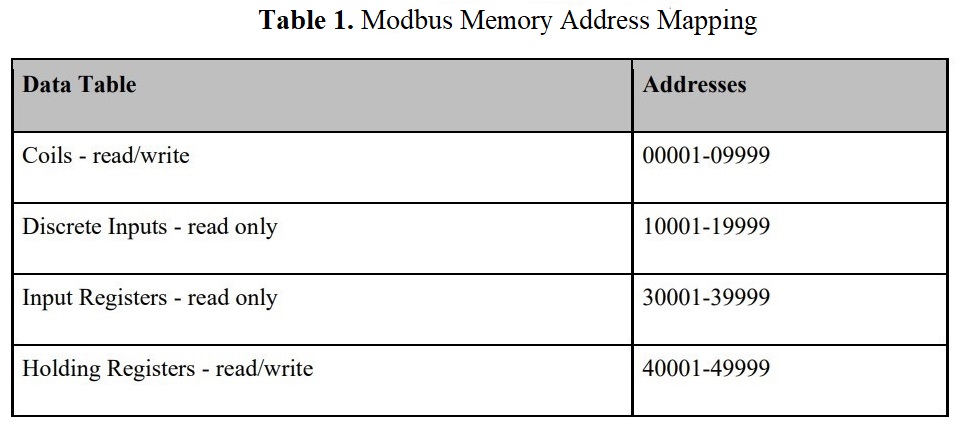}
\end{figure}

Coils and registers are two different categories of storage locations for data. There are two distinct register types for each of these datastore kinds: read/write and read-only. These datastore kinds are all references to specific locations in memory.
\begin{itemize}
    \item a coil is used for storing simple booleans (1 bit). It is read/write and starts from 00001 to 09999;
    \item a discrete input is a read-only type for booleans, starting from 10001 to 19999;
    \item an input register is a read-only type for longer values (16 bits), starting from 30001 to 39999;
    \item a holding register is a read/write type for longer values (16 bits), starting from 40001 to 49999;
\end{itemize}

In the Modbus protocol, the first 20000 bytes of memory are set aside for manufacturer-specific purposes. Data particular to a given device, including setup options or history data, may be stored in this way.

\subsubsection{Manipulating the resource data}
Employing the Metasploit attack module, we discovered all Modbus Slaves existed in the LAN and those hidden behind a Modbus Gateway. By the use of the Metasploit "Modbus client" attack module, we were able to read/write registers/coils on a given Modbus Slave in Fig. 8.  

\begin{figure}[h!]
\centering
\includegraphics[width=1\columnwidth]{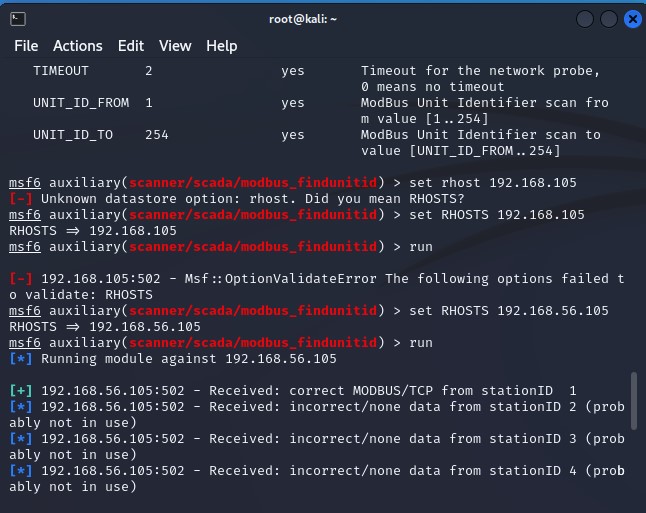}
\caption{The use of Metasploit on Modbus.}
\label{Fig. 7}
\end{figure}

\section{Result and Discussion}
\subsection{Simulation and analytical results}
For this proposed testbed simulation three computers were set up as Master, PLC, and attacker workstations in a lab. Modbus TCP software was used by the Master and PLC workstations to manage coils, and Ettercap was installed on the attacker's workstation to conduct a MiTM attack. Wireshark was used to show how the Master and a PLC workstation communicate. In the initial situation, the Master workstation transmitted a Modbus TCP order to the PLC workstation to switch on a single coil. In the second case, the identical objective was accomplished, but a MiTM attack using an Ettercap filter was added to change the write coil command from the Master workstation to the PLC workstation from on to off. Wireshark displayed the anticipated Modbus TCP traffic between the Master and workstations in scenario one. In case number two, Wireshark identified anomalies in the Modbus TCP interactions, including repeated acknowledgments, unexpected MAC addresses, and retransmission failures. Attackers will continue to choose the easiest route and target the network vulnerabilities with the Modbus TCP protocol as long as an ICS network uses it. Defenders will be able to easily set up a lab to replicate and observe attacks on ICS networks using the research presented in this paper. 

\subsection{Modbus Slave Data Access/Modify}
The MiTM attack on the Modbus TCP communications and regular Modbus TCP communications were both clearly distinguished by Wireshark. Normal connections did not contain TCP faults and contained MAC addresses corresponding to the right workstations. As the MAC address of the Kali Linux workstation was impersonating both the Master and PLC workstations, the MiTM attack, on the other hand, demonstrated that the communications were not typical of Modbus TCP connections. The MiTM attack also showed duplicate acknowledgments and TCP retransmission issues.Additionally, Wireshark illustrated how an Ettercap filter MiTM attack successfully adjusted a request to the PLC from the Master to turn off a coil rather than turn it on. There was no hint given to the Master that the original Modbus TCP command had changed.

Fig. 9. showed the values of the altered registers by the interceptor in the middle from Modbus client to server. By the use of Metasploit, the MiTM attacker has become successful in intercepting the registers and coil values and manipulates those data intentionally to make the system cripple and even in the worst case it may result in system failure or shutdown.

\begin{figure}[h!]
\centering
\includegraphics[width=1\columnwidth]{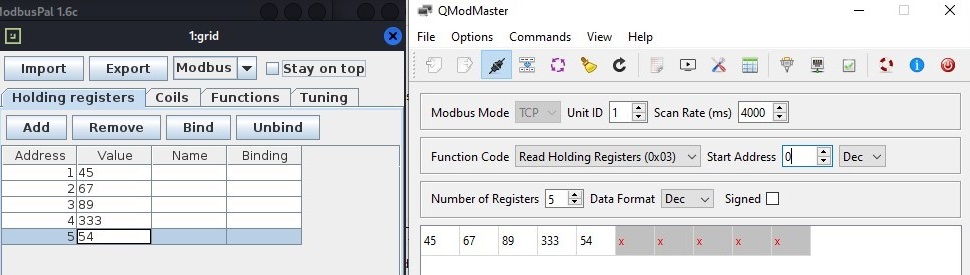}
\caption{The altered registers values by the interceptor in the middle from Modbus client to server.}
\label{Fig. 8}
\end{figure}

\subsection{Analysis packet in Wireshark}
Our assumption is the attacker
is an insider with extensive knowledge of the network and has access to the network connected to the smart grid, or the attacker has physical access to the field devices. Wireshark displayed the anticipated Modbus TCP traffic between the Master and workstations in scenario one. In case number two, Wireshark identified anomalies in the Modbus TCP interactions, including repeated acknowledgments, unexpected MAC addresses, and retransmission failures. Attackers will continue to choose the easiest route and target the network vulnerabilities with the Modbus TCP protocol as long as an SG network uses it. Defenders will be able to easily set up a lab to replicate and observe attacks on SG networks using the Metasploit. As a result of the attack, there has been malicious changes over the registers and coils adversely that impact the system configuration as well as data communication. Hence system disruption is inevitable. To analyze further here is the packet analysis in Wireshark in Fig. 10. displayed the warning message of the system that has gone through the MiTM attack in its Modbus protocol. The message is clearly shown the connection reset action which means the current service is not available more. This is enough to get suspended during the entire process of the simulated testbed of smart grid.

\begin{figure}[h!]
\centering
\includegraphics[width=1\columnwidth]{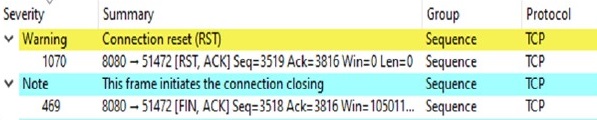}
\caption{The warning message captured in Wireshark after the attack.}
\label{Fig.13}
\end{figure}

\subsection{Limitation and Contribution}
To the best of our knowledge, there hasn’t been much-published research on the testbed-based model to analyze cyber-attacks in smart grids. Although our approach is to develop a testbed that mimics a real-time smart grid system and analyze the effects of the cyber attack on Modbus, it needs to incorporate more hardware into the loop to detect the attack in a different layer of the network. The assessment was carried out only on small scale but needs to validate on a large scale too. In our future endeavors, we plan to perform more potential attacks into other protocols too so that we can observe how the system behaves under or post-attack time. In order to validate we can employ some machine learning tools also for validating our work in the next phase.

\section{Mitigation Techniques}
 To overcome the Modbus protocol's security issues, this article suggests a new secure version based on the Transport Layer Security protocol. According to experimental results, the proposed method achieves request/response times far within the 16.67 ms period of the power grid's 60 Hz cycle, demonstrating a negligible influence on power grid applications. The literature suggests a variety of attack detection and countermeasure techniques to defend against cyberattacks\cite{R19}. Achieving security requires integrating multiple strategies into a global strategy rather than one solution.

\begin{itemize}

\item Network segmentation, which physically separates critical systems and data from less sensitive ones, reduces the attack surface and mitigates MiTM damage. VLANs, firewalls, and access restrictions can limit network access and attack range.

 \item Secure protocols, network segmentation, SSL/TLS certificate verification, two-factor authentication, and software updates can prevent man-in-the-middle attacks. A man-in-the-middle (MiTM) attack may be less likely if a corporation takes these procedures.
 
 \item Two-factor authentication (2FA), which requires a password plus an SMS or email one-time code, may help prevent MiTM attacks. Malicious actors may find it tougher to steal data by impersonating legitimate users. 

 \item Since developers address bugs in updated software, it's crucial to prevent MiTM attacks. Updating operating systems, browsers, and plugins protects against network vulnerability exploitation.
 
\item Checking the SSL/TLS certificate ensures secure and legitimate communication. Users should verify that the website or service they are connecting to has a valid SSL/TLS certificate from a trusted CA.
    
\item Monitoring network traffic for suspicious activity can help prevent man-in-the-middle (MiTM) attacks. Monitoring software can detect suspicious network activity and alert security experts.
 
\item Certificate pinning prevents MiTM attacks employing fake SSL/TLS certificates. "Certificate pinning," which links SSL/TLS certificates to a public key or set of public keys, makes them harder to falsify.

\item Educating people about MiTM attacks and how to avoid them reduces the likelihood of successful assaults. Training programs may encompass safer communication, password hygiene, and phishing detection.

\end{itemize}

\section{Conclusion}
For network vulnerability studies in the smart grid system, a real-time testbed is built to learn about security flaws in the communication protocols used in smart grid systems. In this paper, the testbed has been employed as a testing approach not only to analyze the network vulnerabilities by performing the MiTM attack but also as a means of quantifying the system behaviour. To analyze and investigate the communication network in depth the testbed is deployed on the simulated environment which is based on the Modbus protocol between master and slave that operate on a smart grid system, employing with virtual machines and ICS software. This simulated attack scenario is able to mimic the attacker-victim exposure in an empirical way which is beneficial for vulnerability analysis as well as investigate the mitigation techniques. Future work involves extending the test bed to
explore more possible cyber attacks in other network protocols used in smart grid systems and enquires to develop self-healing techniques in the system during those attacks.

\end{document}